\documentclass[onecolumn,trackchanges]{aastex631}
\usepackage{graphicx}
\usepackage{amssymb}                    
\usepackage{ulem}
\usepackage{float}
\usepackage{rotating}
\usepackage{enumitem}
\usepackage{rotating}
\usepackage{makecell}
\usepackage{appendix}
\usepackage{fixmath}
\usepackage{color}
\usepackage{multirow}

\newcommand{\x}{\ensuremath{\times}}

\usepackage{times}
\usepackage{amsmath}
\submitjournal{ApJ}

\shortauthors{Athiray et al.}
\shorttitle{Can EUV images constrain high-temperature plasma?}

\newcommand{\magixs}{\textit{MaGIXS}}

\newcommand{\vii}{\,{\sc vii}}
\newcommand{\viii}{\,{\sc viii}}
\newcommand{\ix}{\,{\sc ix}}
\renewcommand{\xi}{\,{\sc xi}}
\newcommand{\xvii}{\,{\sc xvii}}

\newcommand{\xvi}{\,{\sc xvi}}

\begin{document}

\title{Can emission measure distributions derived from extreme-ultraviolet images accurately constrain high temperature plasma?}

\correspondingauthor{P. S. Athiray}
\email{athiray.panchap@nasa.gov}

\author[0000-0002-4454-147X]{P.S. Athiray}
\affiliation{Center for Space Plasma and Aeronomic Research, The University of Alabama in Huntsville, Huntsville, AL, USA}
\affil{NASA Marshall Space Flight Center, ST13, Huntsville, AL 35812}

\author[0000-0002-5608-531X]{Amy R.\ Winebarger}
\affil{NASA Marshall Space Flight Center, ST13, Huntsville, AL 35812}

\begin{abstract}
Measuring the relative amount of high-temperature, low-emission-measure plasma is considered to be a smoking gun observation to constrain the frequency of plasma heating in coronal structures. Often, narrowband, extreme ultraviolet images, such as those obtained by the Atmospheric Imaging Assembly (AIA) on the Solar Dynamics Observatory (SDO), are used to determine the emission measure (EM) distribution, though the sensitivity to high temperature plasma is limited.  Conversely, the soft X-ray wavelength range offers multiple high temperature diagnostics, including emission lines of N\,{\sc vii}, O\,{\sc vii}, O\,{\sc viii}, Fe\,{\sc xvii}, Ne\,{\sc ix}, and Mg\,{\sc xi}, which can provide tight constraints to the high temperature plasma in the log T 6.1 to 6.7 (${\sim}$ 1-5+ MK) range.  The Marshall Grazing Incidence X-ray Spectrometer (MaGIXS), a slitless spectrograph launched on a NASA sounding rocket on July 30, 2021 resolved an X-ray bright point in multiple emission lines in the soft X-ray wavelength range.  Using coordinated  observations of the same X-ray bright point from SDO/AIA, we compare and contrast the EM distributions from the EUV image data, the X-ray spectra, as well as the combined EUV and X-ray data set.  In this paper, we demonstrate that EM distributions from SDO/AIA data alone can overestimate the amount of high temperature (log T $>$ 6.4) plasma in the solar corona by a factor of 3 to 15. Furthermore, we present our effort to cross-calibrate Hinode/XRT response functions by comparing the observed XRT fluxes with the predicted ones from combined \magixs-1 + AIA EM analysis.
\end{abstract}

\keywords{Sun:corona}

\section{Introduction}

The  emission measure (EM) distribution is a useful diagnostic to constrain the frequency of heating events in solar coronal structures including X-ray bright points and active regions (AR). The EM distribution represents the amount of thermal plasma integrated along the line of sight derived as a function of temperature. Typical EMs of non-flaring coronal structures exhibit a broken power law relationship, with  hotward (${\beta}$) and coolward (${\alpha}$) slopes. The slopes can be used as a diagnostic for the heating frequency \citep{warren2011a, winebarger2011, tripathi2011, reep2013, athiray2019}.

The continuous, full Sun, high spatial resolution (${\sim}$1.2"), narrow band extreme ultraviolet (EUV) images from the Atmospheric Imaging Assembly (AIA; \citealt{lemen2011}) onboard Solar Dynamics Observatory (SDO) are commonly used to determine the EM distributions. Six AIA EUV channels are dominated by several iron emission lines of varying ionization states, i.e., 94\,\AA\ (Fe\,{\sc xviii}, Fe\,{\sc x}), 131\,\AA\ (Fe\,{\sc viii}, Fe\,{\sc xx}, Fe\,{\sc xxiii}), 171\,\AA\ (Fe\,{\sc ix}), 193\,\AA\ (Fe{\sc xi}, Fe{\sc xii}, Fe\,{\sc xxiv}), 211\,\AA\ (Fe\,{\sc xiv}), 335\,\AA\ (Fe\,{\sc xvi}), but have contributions from several other emissions lines \cite[see ][]{odwyer2010}.
The three channels sensitive to the highest temperatures (94, 131, and 193\,\AA) exhibit a bi-modal thermal response, which makes it difficult to delineate the relative contribution from hot components from the contributions of cool structures along the same line of sight. Additionally, the high temperature response of the 131 and 193\,\AA\ channels are from spectral lines only expected during solar flares, leaving a single channel (94\,\AA) to constrain the high temperature slopes of typical quiescent coronal structures.

Due to this limitation, the EM distributions derived from AIA images can not be expected to constrain the emission at high temperatures $>$ 5MK \citep{su_2018}. Attempts have been made to determine a more accurate EM distribution by combining other instruments with AIA. Inclusion of broadband X-ray images from the X-ray Telescope (XRT; \citealt{golub2007}) and spectrally superior EUV data from the EUV Imaging Spectrometer (EIS; \citealt{culhane1991}), both on the Hinode spacecraft, are often used in conjunction with AIA data to provide additional constraints to the EM solver in narrowing down the T-EM space. However, \cite{winebarger2012} established that EIS and XRT also exhibit a ``blind spot'' for high temperature, low emission measure plasma, which indicate that even a combined data set may be insensitive to the hotward EM slope, ${\beta}$. Using synthetic EM distributions with a range of ${\alpha}$ and ${\beta}$, \cite{athiray2019} showed that existing space instrumentation cannot precisely determine the slope of the high temperature emission and therefore we need instruments with superior spectroscopic capabilities to provide excellent temperature diagnostics, which are accessible in spectral observations in X-rays.

The  Marshall Grazing incidence X-ray Spectrometer (\magixs) instrument \citep{athiray2019, savage2022, champey2022a} was designed to quantitatively measure the high temperature EM slope by distinctly observing high temperature diagnostic emission lines. During the first sounding rocket flight, \magixs-1, the instrument observed spectrally dispersed soft X-ray images of solar coronal structures, from 8 to 30 \AA\, which has given an excellent opportunity to compare the EM determined using AIA with those derived from X-ray observations. For this comparison, we considered the observation of an X-ray bright point from \magixs -1 as a case study.

In this work, for the first time, we quantify the degree to which EM distributions derived from AIA alone can overestimate high temperature, low-emission-measure plasma and demonstrate how a combined AIA and \magixs-1 data set can well constrain the complete temperature range expected in quiescent coronal structures. Section 2 describes the data sets and instruments. Section 3 discusses the EM analysis of the X-ray bright point using \magixs-1 and AIA data independently, and also presents the first combined EM inversion of \magixs-1 spectroheliogram data with AIA images. In Section 4, we compare the predicted fluxes from different EM solutions and quantify the overestimation of high temperature fluxes using only AIA images. In Section 5 we describe the efforts to cross-calibrate Hinode/XRT with AIA and {\magixs}-1 using the inverted EM solutions.  Finally, Summary and Conclusions are discussed in Section 6.

\section{Data and Data Processing}

\begin{figure}[h!]
\centering
    \includegraphics[width=0.8\textwidth]{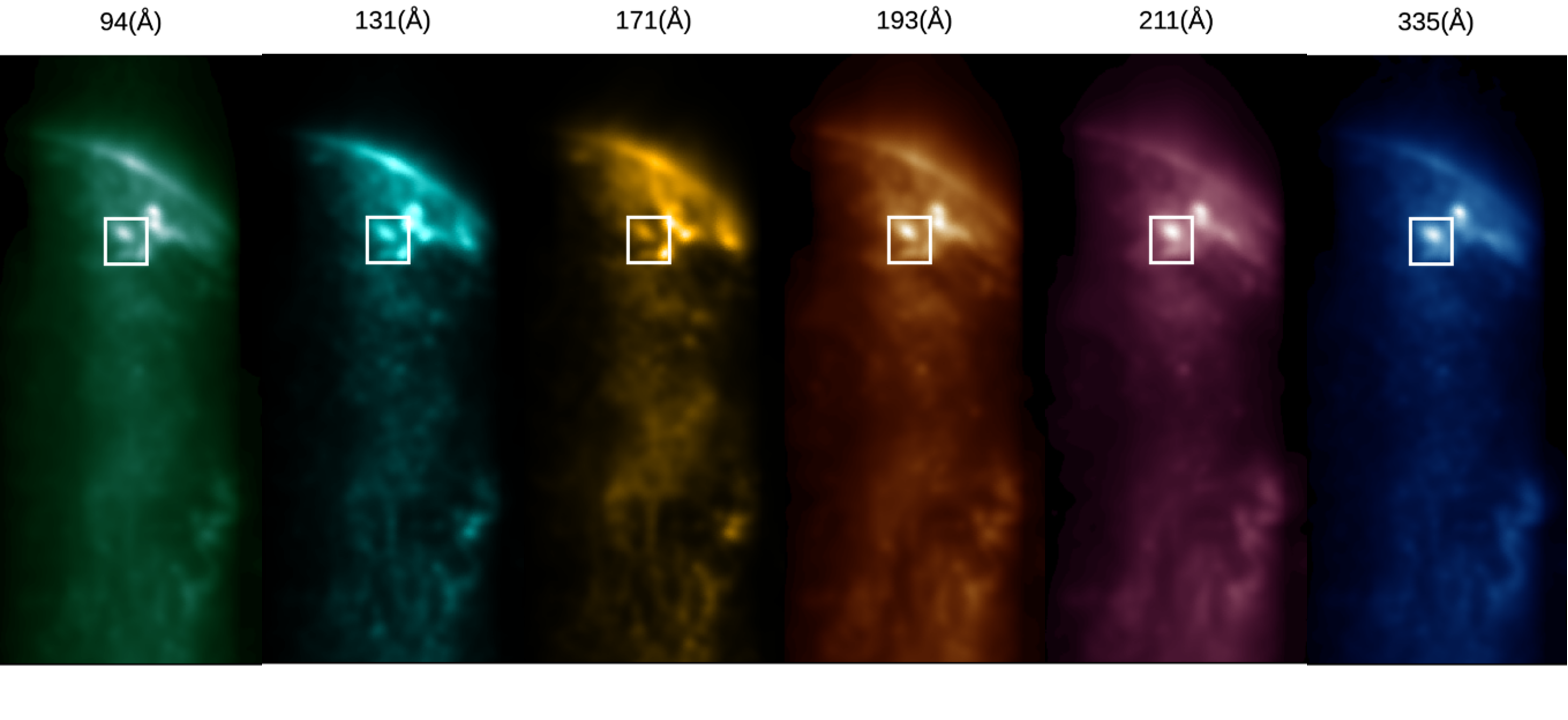}

    \caption{Field of view of \magixs-1 as seen by AIA, where the AIA resolution has been degraded to match \magixs-1. The X-ray bright point used as an example in this study is highlighted with a box.}
    \label{fig:AIA_magixsfov}
\end{figure}

 In this section we give a brief summary of the observations from each instrument and describe the process of data preparation for the analysis.   The \magixs-1 sounding rocket flight occurred on 30 July 2021 from 18:21 to 18:26 UT. The effective \magixs-1 field of view (FOV) covers ${\sim}$ 9.25 \arcmin ${\times}$ 25 \arcmin on the solar disk (see Figure~\ref{fig:AIA_magixsfov}).  Due to its limited observation time and FOV, \magixs-1 observations define the observational target.  The \magixs-1 FOV included two X-ray bright points and a portion of an active region.  In this study, we consider one of the X-ray bright points that was approximately (625\arcsec, 463\arcsec) from Sun center and to the east of Active Region (AR) 12846 highlighted in Figure~\ref{fig:AIA_magixsfov} with a white box.

 The \magixs-1 instrument is a wide-field slot imaging X-ray spectrometer, with a direct focusing Wolter-1 X-ray telescope, two corrective grazing incidence X-ray mirrors, and a reflective grating, which disperses X-rays on to the  CCD \citep{champey2022a}. \magixs-1 observed dispersed X-rays in the wavelength range ${\sim}$ 8\,\AA\ to 30\,\AA\, with 2 second cadence for 296 seconds, during the stable pointing for science observations. \magixs-1 data are spectroheliograms, which contains overlapping dispersed spectra arising from different spatial locations on the Sun.  The spatial plate scale of \magixs-1 is 2.8\arcsec\ in the cross-dispersion direction and varies from ${\sim}$ 5.5\arcsec\ to 9\arcsec\ in the dispersion direction. The Point Spread Function (PSF) of \magixs-1 is ${\sim}$ 27 \arcsec\ \citep{champey2022a}. For this analysis, we utilized \magixs-1 Level 1.5 data products (see \citealt{savage2022} for a description of the data processing).  We average the data over the stable pointing period of science observations of the rocket flight. The \magixs-1 observations are modulated by the instrument vignetting function described in \cite{savage2022}. The vignetting function peaks at the center FOV and gradually decreases on both sides of the field, altering the apparent brightness of the observed features. As described in the \magixs-1 mission paper, the \magixs-1 data have been co-aligned with AIA images using cross-correlation method to determine the absolute pointing.

 SDO/AIA was operational during the \magixs-1 flight, observing the full Sun with all EUV passbands. The data have ${\sim}$ 1.2\arcsec\ spatial resolution (0.6\arcsec\ plate scale) and 12~s temporal cadence. In this analysis, we average the AIA EUV images taken from 18:21 UT to 18:28 UT on 30 July 2021 to match the \magixs\ rocket flight observation. We obtained the cutouts of the full-disk images via the Stanford Joint Science Operations Center (JSOC) Science Data Processing (SDP) center. We selected the \magixs-1 field of view with appropriate 23$^{\circ}$ roll angle to match \magixs-1 observations. We follow standard AIA routines to co-align datacubes onto a common platescale via {\it aia\_prep.pro} distributed in SSW.  We rebinned the AIA pixels to  2.8\arcsec/pix plate scale, while preserving the average intensity per pixel, so as to use the standard AIA temperature response functions.  We also multiplied the EUV images by the \magixs\ vignetting function and finally, the AIA data is convolved with the \magixs-1 PSF.  Figure \ref{fig:AIA_magixsfov} shows the time  averaged cutout images of the AIA channels in the \magixs-1 field of view with the vignetting function and PSF applied.   The X-ray bright point that is used as an example in this paper is highlighted with a rectangle box.

\section{Emission Measure Analysis}

The goal of this research is to quantify the differences in the EM distributions derived from using AIA data alone, \magixs-1 data alone, and a combined \magixs-1 and AIA data set.   To complete this goal, we use the X-ray bright point observed by both \magixs-1 and AIA. We use six AIA passband data viz 94 \AA, 131 \AA, 171 \AA, 193 \AA, 211 \AA, and 335\AA\, (see Figure \ref{fig:AIA_magixsfov}) for the EM analysis. Further, we note that no signature of bursty impulsive heating events such as brightening and/or microflares are observed from this bright point, and the emission is very steady during this time of observation, consistent with high frequency heating \citep[][]{savage2022}.

\subsection{SDO/AIA alone}
To determine the EM distribution of the observed plasma using AIA data alone, we first find the emission measure distribution in the pixels within the region of interest (rectangle box) in Figure \ref{fig:AIA_magixsfov}.  Determination of EM from a set of spectrally impure narrowband AIA observations is an under-determined problem. For this analysis we use the ``standard'' EM solver distributed by the AIA team, {\it aia\_sparse\_em\_init.pro}, in the AIA software with its standard settings \citep{2015ApJ...807..143C} and predict high temperature diagnostic \magixs-1 line intensities and compare with measurements to determine the extent of under/overestimation.

The standard AIA EM technique relies on the assumption that the EM distribution is a linear sum of a series of basis functions, which are Gaussian functions in log T with varying widths.  The routine then finds a sparse EM solution such that the basis function multiplied by the coefficients and convolved with the temperature response functions minimize the difference in the observed and modeled intensities.  Because the EM distribution is assumed to be a sum of Gaussian functions and the inversion algorithm favors sparse solution, the EM distribution is optimized for both sparse and smooth solutions.

The temperature response functions for the SDO/AIA channels are generated using {\it aia\_get\_response.pro} with flags timedepend\_date, eve\_norm, and chianti-fix, using CHIANTI database v 10.1 \citep{Dere2023}. The errors on the AIA intensities are estimated from the function {\it aia\_bp\_estimate\_error.pro} provided by the SDO package in the SSW, which takes a number of instrumental effects into account. The generated response functions have coronal abundances \citep{feldman1992}, ionization equilibrium, and assumed Maxwellian thermal electron distribution.  The inversion is carried ou using default basis widths (${\sigma}$) of 0.1, 0.2 and 0.6 in log T.

\begin{figure}[h!]
     \includegraphics[width=0.5\textwidth]{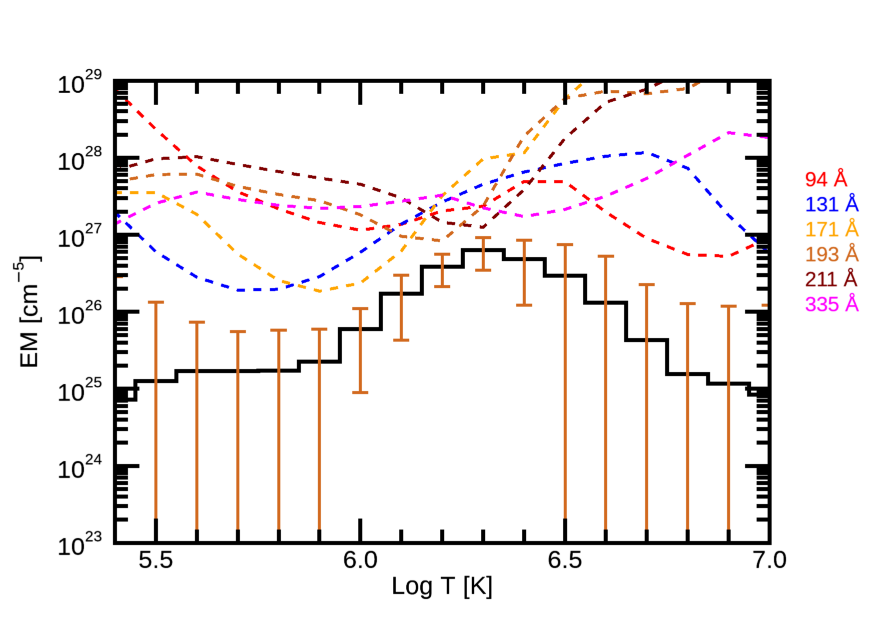}
    \includegraphics[width=0.5\textwidth]{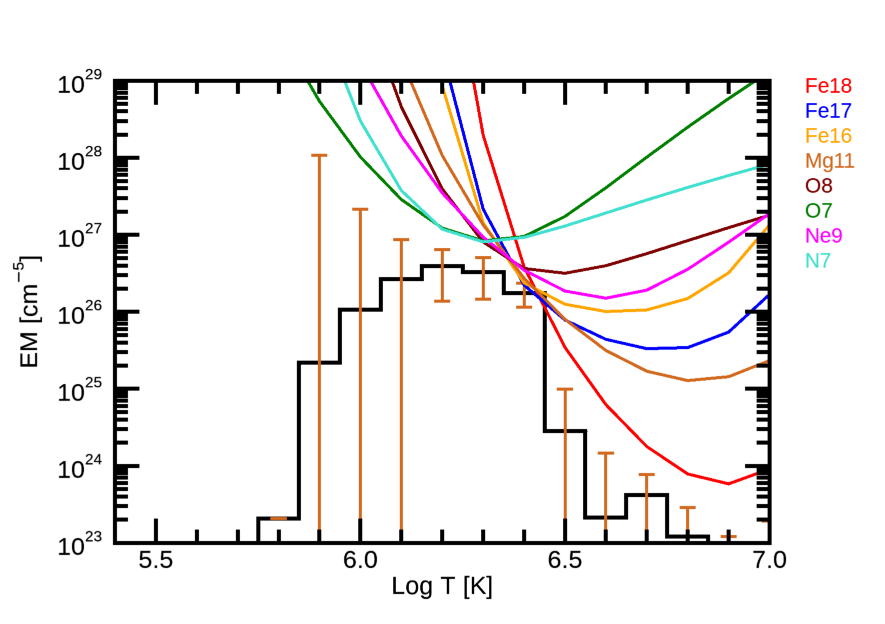}
\caption{The EM solutions obtained for the X-ray bright point using AIA data (left) and \magixs-1 data (right). The best solution for the observed fluxes is shown as a solid black line; The error bars represent change of EM at each temperature bin while the intensities in any of the AIA channel or \magixs\ -1 line is not changed by more than 20\% of the best EM solution (see text for additional details). The EM loci curves of the respective channels or spectral lines are overplotted in different colors.}
    \label{fig:aia_magixs_dem}
\end{figure}

The resulting EM distribution and the corresponding AIA EM loci curves are shown in the left panel of Figure \ref{fig:aia_magixs_dem}. The EM distribution peaks at log T = 6.3 and falls off gradually toward higher temperatures.  Note the EM Loci curves indicate that the AIA 94\,\AA\ channel is the only constraint of the EM distribution at high temperatures (log T $>$ 6.6).

We assess the fidelity of the best EM solution by adopting a similar  approach from \cite{winebarger2012}. We calculate how much emission measure can be added or removed in each temperature bin without varying the intensities in any AIA channel by more than 20\%, typical calibration uncertainty for AIA filter response functions \citep{guennou2013}, and larger than the statistical uncertainty in the average values.

The result is shown as error bars on the EM curve in the left panel of Figure \ref{fig:aia_magixs_dem}.  These error bars show the emission measure curve is well constrained in the temperature range 6.0 ${\leq}$ log T ${\leq}$ 6.4, meaning small changes in the emission measure in those bins result in a change in the AIA intensity in at least one channel that would be larger than the uncertainty in the calibration. The emission measure is poorly constrained outside this range, indicating the solution can vary significantly and have little impact on the intensities. We mention the error bars shown here do not represent statistical error and/or other systematic uncertainties. Instead, they signify the collective sensitivity of AIA channels to individual temperature bins at which an appreciable change in the AIA intensities can be observed.

\subsection{\magixs-1 alone }

Next, we calculate the EM distribution using only \magixs-1 data with the method described in \cite{savage2022} and following the general procedures described in
\cite{Cheung2019} and \cite{winebarger2019}.  We first cast the problem as a set of linear equations, namely
\begin{equation}
    y = Mx
    \label{eqn:lin}
\end{equation}
where $y$ is a one-dimensional array that contains a single row of the \magixs-1 flight data, $x$ is a one-dimensional array of emission measures at different solar locations and temperatures, and $M$ is a matrix that maps emission from each solar location and temperature into the detector.  The \magixs-1 response matrix, $M$, is generated  using the wavelength calibration as a function of field angle determined from flight data and effective area measured pre-flight \citep{Athiray2021}.  Using the CHIANTI atomic database v 10.1  \citep{Dere2023}, (with additional Fe\,{\sc \xvi} lines - see updates below), we construct an isothermal, unit EM instrument response functions with coronal abundances, Maxwellian electron distribution, and an assumption of ionization equilibrium.  These assumptions were based on the best fit solutions given in \cite{savage2022}.

Solving Equation~\ref{eqn:lin} for spectroheliogram is challenging as $y$ contains large number of features, which are highly correlated. We use the Elastic Net regularization technique \citep[][]{zou2005}, a commonly used approach in machine learning to solve linear regression problems, available in Scikit learn via Python library.  It combines two types of regularization namely the sparsity and smoothness while finding convergence to the best solution, which is defined by the equation:

\begin{equation}
    x^{\#} = argmin\left[ ||W(y-Mx)||^2_{2} + {\alpha}{\rho}||x||_{1} + 0.5{\alpha}(1-{\rho})||x||^2_2\right]
    \label{eq:soln}
\end{equation}
where ${\alpha}$ and ${\rho}$ are hyper parameters.  ${\alpha}$ is the magnitude of the penalty term and ${\rho}$ is varied from 0 to 1 to adjust whether the solution is smoother (${\rho}{\rightarrow}0$) or sparser (${\rho}{\rightarrow}1$).   The first term in Equation \ref{eq:soln} is the standard weighted least squares term that minimizes the difference between the observations and the forward calculated observations. The weights, $W$, is a diagonal matrix where the values are $1/{\sigma}$ and ${\sigma}$ is the quadrature sum of read and photon noise.
The second term in Equation \ref{eq:soln} is the $L_1$ norm of $x$, minimizing this term favors a sparse solution.  The third term is the $L_2$ norm of $x$, minimizing this term favors a smooth solution.  Another free parameter is the spatial resolution of the inverted solution, ${\delta}{\theta}$.   We have carried out a systematic study of ${\alpha}$, ${\rho}$, and ${\delta}{\theta}$ and find that a wide range of parameters yield near identical solutions. Results and inferences from the systematic study are beyond the scope of this work and will be reported elsewhere (Athiray et al., in prep). The solutions shown here use ${\alpha}= 5 {\times}10^{-5}$,  ${\rho}= 0.1$, and ${\delta}{\theta}= 8.4$\arcsec.  As mentioned above, the spatial plate scale of \magixs-1 in the dispersion direction varies as a function of wavelength from  ${\sim}5.5$ to $9$ \arcsec / pixel, while the spatial resolution is ${\sim}27$\arcsec\ \citep[see][]{champey2022a}. We find the resolution of the inversion, ${\delta}{\theta}$ to be a roughly a factor of 3 smaller than the point spread function is optimal (Athiray et al, in prep).

We highlight the two updates to the inversion of \magixs-1 spectroheliogram data compared to the description given in \cite{savage2022}. The first update is for the response function, $M$, which incorporates some new atomic calculations pertaining to the transitions of Fe\,{\sc \xvi}. \cite{savage2022} reported an excess emission in \magixs-1 observation compared to the inversion near 15 \AA\, and hypothesized it was due to the missing atomic transitions corresponding to satellite emission lines in that wavelength range. Based on \magixs-1 team request, additional transitions for Fe\,{\sc \xvi} have been provided to the team to complete this updated inversion and will be added to the CHIANTI database at its next release (Giulio Del Zanna, private communication). In this paper, we include the  updated Fe\,{\sc \xvi} transitions in the \magixs-1 response functions.  This inclusion significantly improved the spectral fits to the flight data. A paper summarizing the results of new line identification from \magixs-1 is under preparation. The second update is the addition of weights to the inversion method (see Equation \ref{eq:soln}), which was not employed in the mission paper. The addition of weights improve the inversion and removes minor artifacts arising due to spatio-spectral confusion (see Athiray et al. for additional details).

We average the returned EM distribution over the bright point in question.  We show the average EM distribution along with the EM loci plots for the key \magixs-1 spectral lines in the right panel of Figure \ref{fig:aia_magixs_dem}.  Compared to the EM distribution calculated using AIA alone (left), the amount of high temperature plasma is significantly less and much more constrained by the \magixs-1 spectral lines. However,  because \magixs-1 has no low temperature sensitivity, the shape of the EM distribution is dominated by the smooth/sparse requirements of the solution, and not by observations, for log T $<$ 6.1.

Similar to AIA EM analysis, we vary the EM in every temperature bin and determine how much emission measure can be added or removed without varying the spectrally pure intensities by more than 20\%, the stated  calibration  uncertainty of \magixs\ -1 \citep[][]{Athiray2020,Athiray2021}. We show this variation in emission measure as error bars in Figure~\ref{fig:aia_magixs_dem} (right). The emission measure is well constrained at log T $\ge$ 6.2. Note that error bars are omitted at temperatures less than log T = 5.8 because there are no constraints at these temperatures, meaning the error bars span the range of emission measure shown in the plot.

\subsection{\magixs-1 and AIA}

\begin{figure}[h!]
    \centering
    \includegraphics[width=0.99\textwidth]{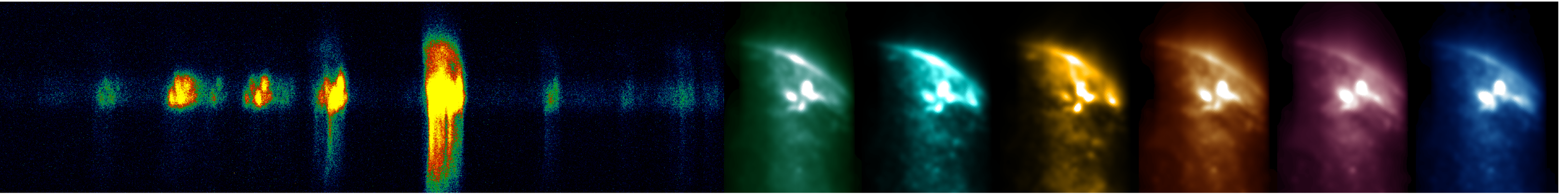}
    \caption{Combined \magixs-1 and AIA coordinated data used for joint EM analysis}
    \label{fig:aia_magixs_combinedfield}
\end{figure}

Finally, we combine the \magixs-1 spectroheliogram data with narrowband EUV images from AIA to perform a joint EM inversion for the first time.
Figure \ref{fig:aia_magixs_combinedfield} shows the combined \magixs-1 and AIA data prepared for a joint EM inversion.

\begin{figure}[h!]
     \centering
    \includegraphics[width=0.5\textwidth]{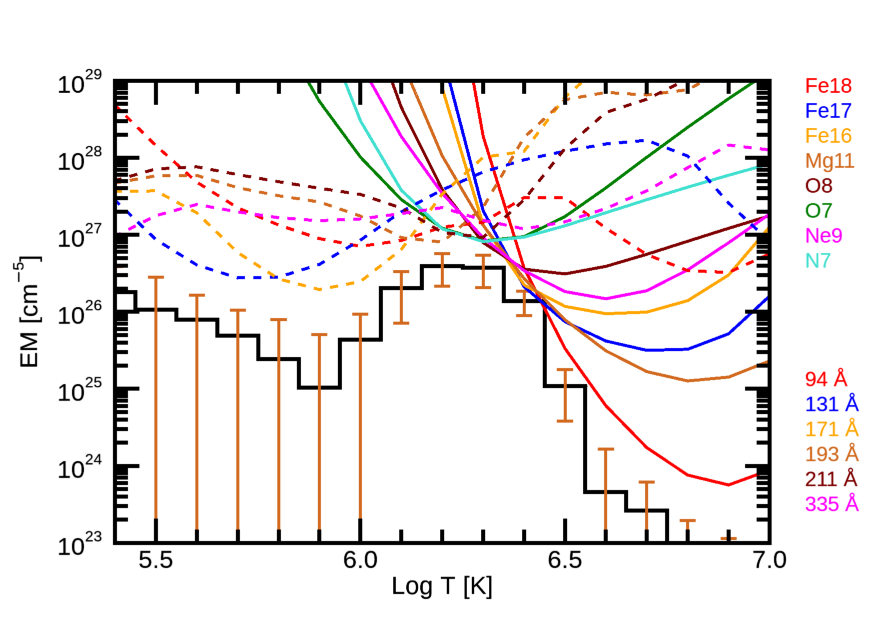}
    \caption{The EM distribution derived using a combined \magixs-1 and AIA data set.  The best solution that matches the observed fluxes in \magixs-1 and AIA is shown as a solid black line; Error bars are determined in the same way as AIA and \magixs-1 analysis (see text). Overplotted are the EM loci curves of \magixs-1 lines (solid colors) and AIA channels (dashed colors).}

    \label{fig:magixsaia_dem}
\end{figure}

The average EM of the X-ray bright point determined using \magixs-1 and AIA data is shown in Figure~\ref{fig:magixsaia_dem}.  When combined, \magixs-1 data requires a much steeper fall off of the EM distribution at temperatures log T ${\geq}$ 6.4 compared to the AIA alone inversion.  Similarly, the amount of low temperature EM increases from both the AIA alone and \magixs-1 + AIA solutions.  Because \magixs-1 is strictly limiting the amount of high temperature plasma, the EM at low temperatures has to increase to account for the intensities in the bi-modal channels, while in the AIA alone EM solution, some of the intensities in the bi-modal channels are accounted for with high temperature emission measure.  Like the previous calculations, we vary the EM in every temperature bin independently to observe a 20\% intensity change in any of the channels/lines used to determine the error bars shown in  Figure~\ref{fig:magixsaia_dem}.

\section{Quantifying the overestimation of high temperature intensities}

In this section, we calculate the average spectral intensities of \magixs-1 lines for the spatially averaged X-ray bright point using the inverted EM solutions. We compare the predicted fluxes determined from inverting only AIA data, only \magixs-1 data, and combined \magixs-1 and AIA data. The predicted fluxes are determined by folding the respective EM distributions through all the spectral line contribution functions obtained using CHIANTI database.

\begin{deluxetable}{cc|ccc|cc}[h!]
\tabletypesize{\small}
\tablewidth{6in}
\tablecolumns{3}
\tablecaption{Average line intensities in some of key \magixs-1 spectral lines predicted using different EM solutions \label{tab:magixs_line_intensities}}
\tablehead{\colhead{Spectral }& \colhead{Temperature} & \multicolumn{3}{c}{Predicted intensity using EM solutions} & \multicolumn{2}{c}{Ratios}\\
\colhead{}& \colhead{} & \multicolumn{3}{c}{${\times}10^{10}$ Ph/s/sr/cm$^2$} & \colhead{AIA /} & \colhead{\magixs-1 /}\\
\colhead{Line [\AA]}& \colhead{[log T]} & \colhead{AIA} & \colhead{\magixs-1} & \colhead{\magixs-1 + AIA}  & \colhead{\magixs-1 + AIA} & \colhead{\magixs-1 + AIA}}
\startdata
N \vii\ 24.770& 6.30 & 4.10& 2.05& 2.07&  1.98& 0.99\\
O \vii\ 21.602& 6.30& 34.40&18.49&18.42& 1.87& 1.00\\
O \vii\ 21.801& 6.30& 6.28& 3.63& 3.61&  1.74& 1.01\\
O \vii\ 22.101& 6.30& 25.46&14.65&14.54&  1.75& 1.01\\
O \vii\ 18.627& 6.30& 4.22& 2.11& 2.11&  2.00& 1.00\\
\hline
O \viii\ 18.967& 6.50& 19.62& 5.47& 5.36&  3.66& 1.02\\
O \viii\ 16.006& 6.50& 2.08& 0.53& 0.51&  4.08& 1.04\\
Ne \ix\ 13.447& 6.60& 2.36& 0.48& 0.47&  5.02& 1.02\\
Ne \ix\ 13.553& 6.60& 0.49& 0.11& 0.11&  4.45& 1.00\\
Ne \ix\ 13.699& 6.60& 1.68& 0.38& 0.38&  4.42& 1.00\\
Fe \xvi\ 15.211& 6.60& 1.73& 0.26& 0.24&  7.21& 1.08\\
\hline
Mg {\xi}\ 9.169& 6.80& 0.62& 0.04& 0.04&  15.5& 1.00\\
Fe \xvii\ 15.013&6.80& 13.53& 1.02& 0.98& 13.81& 1.04\\
Fe \xvii\ 15.262&6.80 & 4.01& 0.31& 0.30&  13.37& 1.03\\
Fe \xvii\ 15.453& 6.80& 0.90& 0.08& 0.08&  11.25& 1.00\\
Fe \xvii\ 16.776& 6.80& 10.64& 0.95& 0.90&  11.82& 1.06\\
Fe \xvii\ 17.051& 6.80& 14.25& 1.29& 1.22&  11.68& 1.06\\
\hline
\enddata
\end{deluxetable}
Table \ref{tab:magixs_line_intensities} lists the average spectral intensities in units of ${\times}10^{10}$ Ph/s/sr/cm$^2$. Column 1 and 2 gives the ion name, its rest wavelength, and the log T where the emissivity function peaks. We group the spectral lines into three different temperature ranges, log T=6.3, 6.5${\leq}$ log T ${\leq}$ 6.6, and log T = 6.8. Columns 3, 4, and 5 gives the spectral line fluxes calculated from the EM solutions determined from AIA data only, \magixs-1 data only, and combined \magixs-1 + AIA data. Here, we consider the \magixs-1 + AIA EM solution to be the best representation of the plasma temperature distribution.  Finally, we then find the intensity ratios for AIA and \magixs-1 to the \magixs-1 + AIA solution, which are listed in columns 6 and 7.   A ratio value departing from unity, and above 20\%, is considered to over-predict (or) under-predict the  intensities with respect to the  intensities determined from the combined EM solution.

The ratio of the intensities calculated from AIA data alone to the \magixs-1 + AIA solution (column 6) clearly shows a systematic overestimation of warm and hot emission lines. We observe a gradual increase in the extent of overestimation for an increase in peak emissivity temperature. For instance, emission lines with peak emissivity temperature, log T = 6.30 (N\,{\sc vii}, O\,{\sc vii}), are overestimated by a factor ${\sim}$ 2; emission lines with peak emissivity temperature of 6.50 ${\leq}$ log T ${\leq}$ 6.60 (O\,{\sc viii}, Ne\,{\sc ix}, and Fe\,{\sc xvi}) are overestimated by a factor ${\sim}$ 3 to 7; emission lines  with peak emissivity temperature at  log T $=$ 6.80 (Mg\,{\sc xi}, Fe\,{\sc xvii}) are overestimated by a factor ${\sim}{\times}$ 11 to 15. However, the ratio of \magixs-1 alone data with \magixs-1 + AIA data did not show a significant change in the predicted fluxes. This result demonstrates that AIA channels do not offer tight constraints to the plasma emission greater than log T = 6.3, at which spectral lines from \magixs-1 are chosen to precisely determine the low emission plasma.

\section{Cross-calibration with Hinode/XRT}
Here, we investigate the compatibility of the derived EM solutions with the Hinode/XRT observations by computing the predicted XRT intensities and comparing them with the measured ones. However, during the \magixs-1 flight observation, XRT was targeted to observe the active region AR 12849 near the south west portion on the disk, hence there is no coordinated observation available for the bright point under study.  Therefore, we consider a nearest time available data from XRT synoptic data archive, which was ${\sim}$ 30 minutes prior to \magixs-1 observation time, with an assumption that the bright point did not evolve significantly within this duration.  We include the Thin-Be/Open, Al-Mesh/Open, and Al-Poly/Open filter combinations with exposure times 23.14\,sec, 2.90\,sec, and 4.09\,sec respectively in this study. We follow standard XRT data processing methods and determine the average intensity of the X-ray bright point, which is given in Table~\ref{tab:xrt_line_intensities} (Column 2).

\begin{deluxetable}{|c|c|ccc|}[h!]
\tabletypesize{\small}
\tablewidth{6in}
\tablecolumns{3}
\tablecaption{Comparison of the measured XRT intensities for the X-ray bright point with the values predicted using different EM solutions. \label{tab:xrtflux}}
\tablehead{\colhead{Hinode/XRT}& \colhead{Measured Intensity} & \multicolumn{3}{c}{Predicted Intensity using EM solutions (DN/s)}\\
\colhead{Channel}& \colhead{(DN/s)} &\colhead{AIA alone} & \colhead{\magixs-1 alone} & \colhead{\magixs-1 + AIA}
\label{tab:xrt_line_intensities}}
\startdata
XRT - Be thin & 4.7 & 18.8 & 2.1 & 2.0 \\
XRT - Al Mesh & 88.9 & 126.0  & 39.0 & 37.8  \\
XRT - Al Poly & 58.6 & 110.0 & 24.6 & 24.0 \\
\hline
\enddata
\end{deluxetable}

Our aim here is to predict the average XRT intensity for the bright point using different EM solutions and compare it with the measured intensity from the three filter combinations. Particularly, the joint EM solution with \magixs-1 and AIA gives an excellent opportunity to cross-calibrate the XRT response, which has a good overlap in the temperature sensitivity. We calculate the predicted intensities for different EM solutions by folding EM through the standard XRT filter temperature response functions obtained using solar soft routine {\it make\_xrt\_temp\_resp.pro} using appropriate time dependent contamination, which are given in Table \ref{tab:aia_line_intensities} (Column 3, 4 and 5). We observe that AIA EM solution systematically over predicts XRT intensity by a factor of 1.4-4, whereas \magixs-1 and \magixs-1 + AIA solutions systematically under predict XRT intensity. Interestingly, the predicted intensities for all the 3 XRT filters using \magixs-1 alone and \magixs-1 + AIA EM solutions are consistently smaller by a factor ${\sim}$ 2.0.  We interpret the excess emission predicted by AIA alone solution is due to the fact that AIA channel response functions is insensitive to high temperature emission, log T${\geq}$ 6.5 (see \citealt{athiray2019}), where XRT is sensitive. \magixs-1 data are spectrally superior to offer precise high temperature diagnostics. Therefore, predictions of XRT intensities from \magixs-1 alone and \magixs-1 + AIA EM would offer unique insights into cross calibration of the instruments. The predicted XRT intensities yield values a factor ${\sim}$ 2 lower than the measured XRT intensities, which would indicate that XRT temperature response functions require a cross-calibration factor (${\sim}$2).

\section{Summary and Discussion}

We have presented the results of a detailed analysis of the observations of an X-ray bright point using \magixs-1 sounding rocket and SDO/AIA data to diagnose the plasma temperature distribution. In this work, we carried out an independent EM analysis using the observations from both the instruments, and also performed a joint EM analysis, for the first time, by combining \magixs-1 spectroheliogram data with narrowband AIA images. This approach of combined EM analysis with \magixs-1 and AIA data provides an unprecedented temperature coverage 5.4 $<$ log T $<$ 7.0, with \magixs-1 constraining the shape of fall off at high temperatures, while the cooler channels of AIA offer constraints to the emission at lower temperatures, as shown in Figure \ref{fig:magixsaia_dem}.

We use the resultant EM distribution from both the instruments combined to predict the fluxes of high temperature diagnostic \magixs-1 lines and compare the results with the EM distribution based on \magixs-1 and AIA alone.  The comparisons are summarized in Table \ref{tab:magixs_line_intensities}, which shows that the EM distribution using AIA data alone overestimates the line fluxes. Furthermore, the extent of overestimation increases with the peak emissivity temperature of the emission lines. Specifically, AIA alone EM solution over predicts log T = 6.3 by a factor of ${\sim}$2; log T = 6.4 to 6.6 by a factor of ${\sim}$ 3 to 7; log T =  6.8 by a factor of ${\sim}$ 11 to 15. In a sharp contrast, the flux predicted using combined \magixs-1 + AIA EM distribution matches closely (within 20\%) with the results from \magixs-1 alone EM distribution, which is consistent with our expectation that high temperature diagnostic lines are accurately modeled using \magixs-1 data.

\begin{figure}
\centering
    \includegraphics[width=0.5\textwidth]{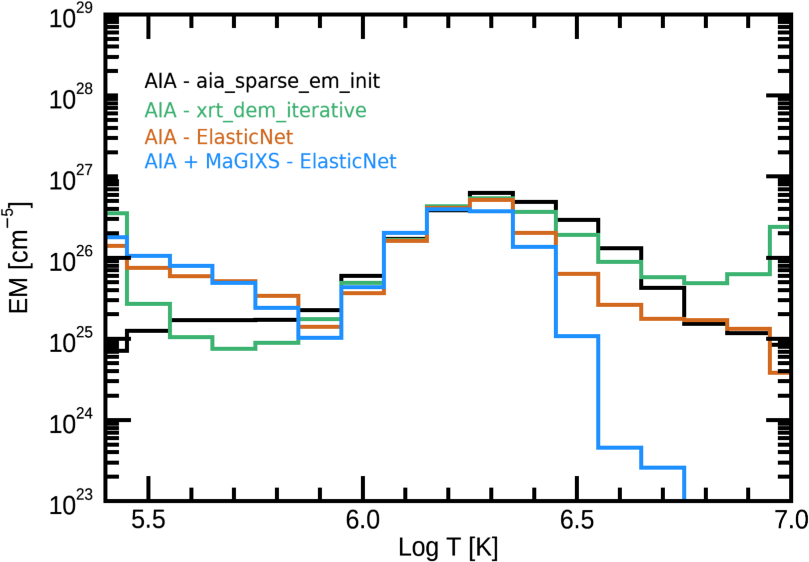}
    \caption{Comparison of EM distribution determined using AIA data only employing different EM solvers viz., {\it aia\_sparse\_em\_init}, {\it xrt\_dem\_iterative}, and ElasticNet, along with the solution using combined MaGIXS-1 + AIA data set. This clearly demonstrates that AIA only solution systematically result in a shallow EM distribution irrespective of the EM solver.}
    \label{fig:EMSolvercmp}
\end{figure}

We emphasize that several EM solver algorithms are available to invert AIA EUV data \citep[e.g.][]{kashyap1998, weber2004, aschwanden2011, hannah2012, plowman2013, 2015ApJ...807..143C} and each method exhibit a set of strengths and also some weaknesses associated in the reconstruction \citep[for instance see][]{aschwanden_bench2015, Massa2023}. The output from any solver results in a plausible solution that matches the observed intensities in all six AIA channels. Therefore, different possible solutions are available to any set of AIA observations. Without additional constraints from high temperature sensitivity instruments like \magixs-1, however, we would expect  excess high temperature emission irrespective of the EM solver employed. To validate this, we performed inversion of AIA data only using three different solvers, viz {\it aia\_sparse\_em\_init} method \citep{2015ApJ...807..143C}, {\it xrt\_dem\_iterative} \citep{golub2004, weber2004} method, and the ElasticNet method \citep[][]{zou2005}. The solutions are compared in Figure \ref{fig:EMSolvercmp}. The first two solvers are the most commonly used and considered as `standard' for EM determination from narrowband images. The ElasticNet method is currently being used for inverting  spectroheliogram data such as \magixs-1. As expected, different solvers resulted in a different solutions. Qualitatively, Figure \ref{fig:EMSolvercmp} shows inverting AIA data using different solvers systematically overestimates the high temperature emission, as compared to the solution determined using combined AIA + \magixs-1 data. The {\it aia\_sparse\_em\_init} and {\it xrt\_dem\_iterative} method solutions yield similar EM curves at high temperature, while the ElasticNet solution found slightly less EM at log T = 6.4 - 6.7. We hereby infer the excess high temperature emission determined by inverting AIA data do not strongly depend on the EM solver method but rather shows the limited sensitivity of AIA channels at these temperatures.

This result is not surprising as the temperature response of AIA channels do not offer sensitivity to high temperatures alone. Instead, the channels that are sensitive to high temperatures, 94, 131, and 193 \AA, exhibit a bimodal response, which means they can detect both high and low temperature plasma. The low temperature emission from  cool structures along the line of sight often dominates the observed signal, except possibly in flares.   Furthermore, the amount of cool plasma along the line of sight limits the sensitivity to  high temperature emission.  For instance, consider the EM loci curve of 94\,\AA\ (red dashed line) in Figure \ref{fig:aia_magixs_dem} (left).  If there were more (less) emission at log T = 6.1, the EM Loci curve would move higher (lower).  The error bars at high temperature would follow this motion, so the uncertainty in the high temperature emission is directly proportional to the amount of lower temperature emission along the line of sight.  This ambiguity can lead to an overestimation of  high temperature emission measure.  The difficulty in constraining high temperature, low emission measure plasma in AIA channels is akin to the blind spot established in Hinode/EIS and XRT by \cite{winebarger2012}.

 As AIA lacks true high temperature diagnostics, \magixs-1 has no sensitivity to plasma with temperatures less than Log T = 6.1.  Hence, we also use the resultant EM distributions to predict fluxes in the observed AIA channels and then compare them with actual AIA measurements (Table \ref{tab:aia_line_intensities}). We mention that we only consider calibration uncertainty (20\%) on the AIA response functions, based on \cite{guennou2013}, which also lists other possible sources of uncertainties. Therefore, we look for flux variations beyond 20\% in the comparison to mark an overestimation or underestimation. We notice that the predicted intensities using \magixs-1 alone solution (column 3) underestimate emission in channels 131 \AA\ (${\sim}$ 52\% less), 211 \AA\ (${\sim}$ 36\% less), and 335 \AA\ (${\sim}$ 42\% less). The 131 \AA\ channel exhibit a bimodal response with sensitivity to lower and higher temperatures from Fe\,{\sc viii, xx, xxiii}. Although \magixs-1 can detect hot emission from Fe \,{\sc xx, xxiii}, which are not observed in the bright point under study, therefore we interpret 131 \AA\ emission is mainly due contributions from Fe\,{\sc viii} ${\sim}$ 0.5\,MK, which \magixs-1 cannot detect. Fluxes predicted from \magixs-1 + AIA (column 4) provides a closer agreement for all the AIA channels, which is consistent with the EM loci plots shown in Figure \ref{fig:magixsaia_dem}.

\begin{deluxetable}{ccccc}[h!]
\tabletypesize{\small}
\tablewidth{6in}
\tablecolumns{3}
\tablecaption{Comparison of the measured average intensities in six AIA channels for the X-ray bright point with the values predicted using different EM solutions in units of DN/s. \label{tab:key_spec_lines}}
\tablehead{\colhead{Channel}& \colhead{Measured Flux} & \multicolumn{3}{c}{Predicted Flux using EM solutions (DN/s)}\\
\colhead{}& \colhead{(DN/s)} &\colhead{AIA alone} & \colhead{\magixs-1 alone} & \colhead{\magixs-1 + AIA}
\label{tab:aia_line_intensities}}
\startdata
AIA - 94 \AA\ & 1.3   & 1.6   & 1.1 & 1.0   \\
AIA - 131 \AA\ & 5.2   & 3.6  & 2.5  & 5.3   \\
AIA - 171 \AA\ & 160.0   & 166.0 & 194.8  & 173.8   \\
AIA - 193 \AA\ & 255.8   & 257.8 & 247.2 & 247.4   \\
AIA - 211 \AA\ & 95.7   & 88.9   &  61.4 & 65.7   \\
AIA - 335 \AA\ & 1.7   &  1.8  & 1.0  &1.3    \\
\hline
\enddata
\end{deluxetable}

\begin{figure}
    \includegraphics[width=0.45\textwidth]{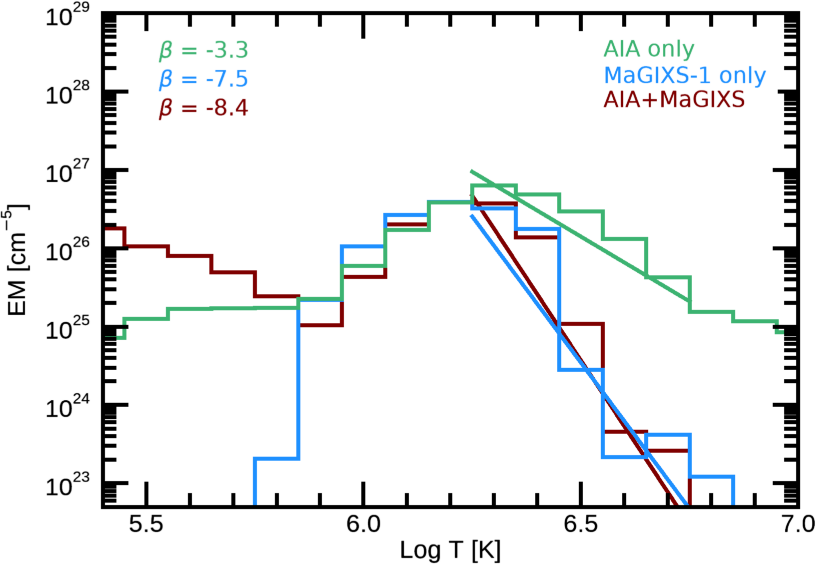}
    \includegraphics[width=0.52\textwidth]{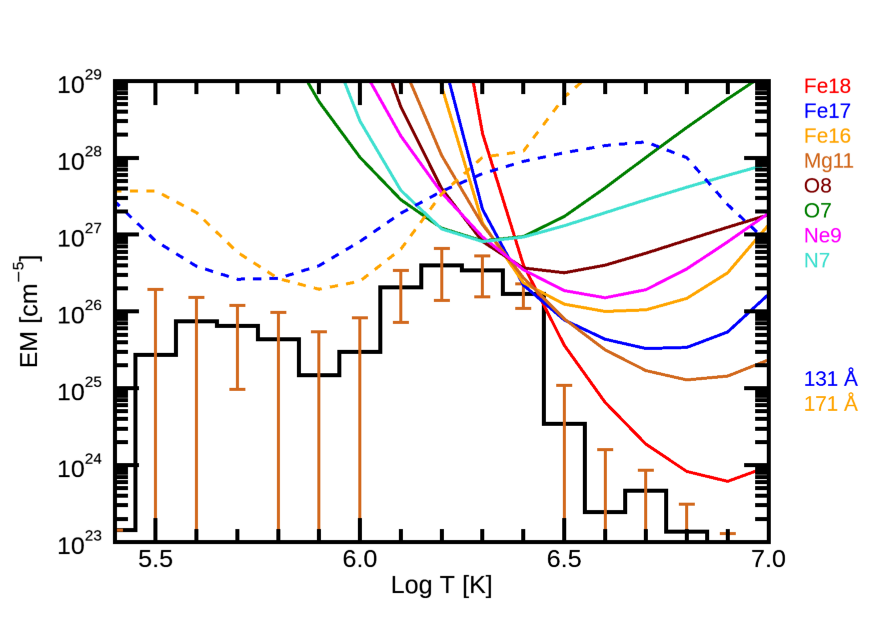}

    \caption{(Left) Comparison of EM distribution derived using AIA data only, \magixs-1 data only, and combined \magixs-1 + AIA data set, along with fits to the high temperature slope (${\beta}$). This clearly demonstrates that AIA only solution results in a shallow  EM distribution compared to \magixs-1 only and \magixs-1 + AIA solution, which results in a steep EM distribution (right). Inverted EM solutions with \magixs-1 and two AIA channels 131 \AA\ and 171 \AA\ along with respective EM loci curves (colored curves). This plot portrays that a wide temperature coverage with an excellent coronal plasma temperature diagnostics can be achieved with high resolution X-ray imaging spectroscopy and EUV imagers.}
    \label{fig:magixs_aia_131_171_dem}
\end{figure}

The slope of the high temperature EM distribution offers an important constraint on the timescale between heating events \citep{athiray2019, barnes2019}. If the heating occurs at a high frequency, with events spaced at intervals much shorter than a cooling time, the EM distribution will have a steep fall off at high temperatures, while low frequency heating will exhibit a shallow hotward slope. Figure \ref{fig:magixs_aia_131_171_dem} (Left) shows a comparison of the EM solutions from AIA alone, \magixs-1 alone, and \magixs-1 + AIA data, along with the hotward slope (${\beta}$) fitted from the peak of the EM distribution. Overestimation of high temperature emission using AIA data alone gives shallower ${\beta}$ compared to the steep ${\beta}$ from the \magixs-1 and \magixs-1 + AIA solutions. This study once again confirms the X-ray bright point observed by \magixs-1 flight must be heated at a relatively high frequency so that the loops are close to equilibrium and the distribution of temperatures is narrow.

The combined EM analysis using \magixs-1 and AIA instrument data allowed us to investigate the cross-calibration of the Hinode/XRT response functions. Using the XRT synoptic data from the nearest time of \magixs-1 observation, we compared the observed XRT intensities in the X-ray bright point understudy against the predicted XRT intensities using the joint EM solution from \magixs-1 + AIA. We find that the predicted XRT intensities is smaller, by a factor of ${\sim}$ 2, than the observed XRT intensities. This implies that XRT response functions require a cross-calibration factor of ${\sim}$ 2. This factor agrees closely with several earlier reported studies from combined EM analysis with high temperature sensitivity X-ray instruments such as NuSTAR, FOXSI \citep{Wright_2017, Athiray2020}. The source for this cross-calibration factor is still unknown. However, we find reports from the earlier years of Hinode mission comparing the EM distributions from the EUV Imaging Spectrometer (EIS) and XRT, which observes this discrepancy. For instance, \cite{kimble2011, Kimble2011PhD} reported EIS - XRT cross-calibration factor using X-ray bright point observations from 2010. The study finds that for a combined EIS - XRT analysis, the observed XRT intensities must be multiplied by a ${\times}$ 0.25, which in other words mean multiply the XRT response functions by a factor of 4. Another study by \cite{Paola2011}, reported EM distributions determined from a coordinated EIS - XRT observation, from a non-flaring active region. The study found that EIS EM solutions are consistently smaller than XRT EM solutions by a factor of ${\sim}$ 2, while exhibit similar width and peak temperature. Although the study eluded a cross-calibration factor for a combined EIS - XRT analysis, however, interpreted this discrepancy due to the influence of elemental abundances (see \cite{Paola2011} for further details). Oddly, no discrepancy between EIS and XRT was found in \cite{winebarger2012}. Also, investigation by \cite{odwyer2014} showed that observed XRT fluxes agree with EIS observations for active regions, and are strongly dependent on the elemental abundances.

We emphasize that \magixs\ -1 is spectrally superior, and a well calibrated instrument \citep{Athiray2020, Athiray2021} to precisely quantify the amount of high temperature plasma, which could be used to cross-calibrate XRT filters due to a good overlap in the temperature sensitivity. However, lack of coordinated XRT observation for the bright point understudy combined with the uncertainty on the vignetting model, strictly limits us to complete the cross-calibration. Nevertheless, this effort strongly motivates the need for cross-calibration of XRT, possibly with the upcoming second flight of \magixs-2, scheduled for summer 2024.

The combined EM analysis using \magixs-1 + AIA also revealed an interesting insight to the design of future instrumentation for a wide range of temperature coverage along with high sensitivity.  Figure \ref{fig:magixsaia_dem} demonstrates that two (131 \AA\ , 171\AA\ )  of the six AIA channels, along with the assumption that the EM curve is smoothly varying, are sufficient to constrain the plasma temperature distribution at temperatures log T ${\leq}$ 6.1. To establish this result, we have performed a combined inversion of \magixs-1 + AIA with only 131 \AA\ and 171 \AA\ channels added, which is shown in Figure \ref{fig:magixs_aia_131_171_dem} (Right). A future instrument design with spatially dispersed spectral images in X-rays combined with 131 and 171 \AA\ EUV images would offer an excellent thermal plasma diagnostics covering a wide range of coronal temperatures.

This approach of combined inversion using spatial-spectral overlapped data and images (EUV or X-ray) will set a precedent for the analysis of future missions such as the Cubesat Imaging X-ray Solar Spectrometer (CubIXSS) \citep{CubIXSS2021} and potential small explorer (currently in Phase A) The EUV CME and Coronal Connectivity Observatory(ECCCO) \citep{kathy2022},  which will carry spectroheliogram instruments along with imagers.

\section{Acknowledgements}
\begin{acknowledgements}
    We acknowledge the Marshall Grazing Incidence X-ray Spectrometer \magixs\ instrument team for making the data available through the 2014 NASA Heliophysics Technology and Instrument Development for Science (HTIDS) Low Cost Access to Space (LCAS) program, funded via grant NNM15AA15C. MSFC/NASA led the mission with partners including the Smithsonian Astrophysical Observatory, the University of Central Lancashire, and the Massachusetts Institute of Technology. The authors thank Mr. Arthur Hochedez for his help and support with Python scripts. \magixs\ was launched from the White Sands Missile Range on 2021 July 30. The authors thank Dr. Katharine Reeves (KR), Project Scientist, Hinode/XRT team for their inputs and discussion. KR is supported by contract NNM07AB07C from NASA to SAO. CHIANTI is a collaborative project involving George Mason University, the University of Michigan (USA), University of Cambridge (UK) and NASA Goddard Space Flight Center (USA).
\end{acknowledgements}
\bibliography{solar_magixs, solar}{}

\begin{thebibliography}{}
\expandafter\ifx\csname natexlab\endcsname\relax\def\natexlab#1{#1}\fi
\providecommand{\url}[1]{\href{#1}{#1}}
\providecommand{\dodoi}[1]{doi:~\href{http://doi.org/#1}{\nolinkurl{#1}}}
\providecommand{\doeprint}[1]{\href{http://ascl.net/#1}{\nolinkurl{http://ascl.net/#1}}}
\providecommand{\doarXiv}[1]{\href{https://arxiv.org/abs/#1}{\nolinkurl{https://arxiv.org/abs/#1}}}

\bibitem[{{Aschwanden} \& {Boerner}(2011)}]{aschwanden2011}
{Aschwanden}, M.~J., \& {Boerner}, P. 2011, \apj, 732, 81,
  \dodoi{10.1088/0004-637X/732/2/81}

\bibitem[{{Aschwanden} {et~al.}(2015){Aschwanden}, {Boerner}, {Caspi},
  {McTiernan}, {Ryan}, \& {Warren}}]{aschwanden2015}
{Aschwanden}, M.~J., {Boerner}, P., {Caspi}, A., {et~al.} 2015, \solphys, 290,
  2733, \dodoi{10.1007/s11207-015-0790-0}

\bibitem[{{Athiray} {et~al.}(2019){Athiray}, {Winebarger}, {Barnes},
  {Bradshaw}, {Savage}, {Warren}, {Kobayashi}, {Champey}, {Golub}, \&
  {Glesener}}]{athiray2019}
{Athiray}, P.~S., {Winebarger}, A.~R., {Barnes}, W.~T., {et~al.} 2019, \apj,
  884, 24, \dodoi{10.3847/1538-4357/ab3eb4}

\bibitem[{{Athiray} {et~al.}(2020){Athiray}, {Winebarger}, {Champey},
  {Kobayashi}, {Vigil}, {Haight}, {Johnson}, {Bethge}, {Rachmeler}, {Savage},
  {Beabout}, {Beabout}, {Hogue}, {Guillory}, {Siler}, {Wright}, \&
  {Kegley}}]{Athiray2020}
{Athiray}, P.~S., {Winebarger}, A.~R., {Champey}, P., {et~al.} 2020, \apj, 905,
  66, \dodoi{10.3847/1538-4357/abc268}

\bibitem[{{Athiray} {et~al.}(2021){Athiray}, {Winebarger}, {Champey},
  {Kobayashi}, {Savage}, {Beabout}, {Beabout}, {Broadway}, {Bruccoleri},
  {Cheimets}, {Golub}, {Gullikson}, {Haight}, {Heilmann}, {Hertz}, {Hogue},
  {Johnson}, {Kegley}, {Kolodziejczak}, {Madsen}, {Schattenburg}, {Siler},
  {Vigil}, \& {Wright}}]{Athiray2021}
---. 2021, \apj, 922, 65, \dodoi{10.3847/1538-4357/ac2367}

\bibitem[{{Barnes} {et~al.}(2019){Barnes}, {Bradshaw}, \& {Viall}}]{barnes2019}
{Barnes}, W.~T., {Bradshaw}, S.~J., \& {Viall}, N.~M. 2019, \apj, 880, 56,
  \dodoi{10.3847/1538-4357/ab290c}

\bibitem[{{Caspi} {et~al.}(2021){Caspi}, {Shih}, {Panchapakesan}, {Warren},
  {Woods}, {Cheung}, {DeForest}, {Klimchuk}, {Laurent}, {Mason}, {Palo},
  {Seaton}, {Steslicki}, {Gburek}, {Sylwester}, {Mrozek}, {Kowali{\'n}ski},
  {Schattenburg}, \& {The CubIXSS Team}}]{CubIXSS2021}
{Caspi}, A., {Shih}, A.~Y., {Panchapakesan}, S., {et~al.} 2021, in American
  Astronomical Society Meeting Abstracts, Vol.~53, American Astronomical
  Society Meeting Abstracts, 216.09

\bibitem[{{Champey} {et~al.}(2022){Champey}, {Winebarger}, {Kobayashi},
  {Athiray}, {Hertz}, {Savage}, {Beabout}, {Beabout}, {Broadway}, {Bruccoleri},
  {Cheimets}, {Davis}, {Duffy}, {Golub}, {Gregory}, {Griffith}, {Haight},
  {Heilmann}, {Hogue}, {Hohl}, {Hyde}, {Kegley}, {Kolodzieczjak}, {Ramsey},
  {Ranganathan}, {Robertson}, {Schattenburg}, {Speegle}, {Vigil}, {Walsh},
  {Weddenorf}, \& {Wright}}]{champey2022a}
{Champey}, P.~R., {Winebarger}, A.~R., {Kobayashi}, K., {et~al.} 2022, Journal
  of Astronomical Instrumentation, 11, 2250010,
  \dodoi{10.1142/S2251171722500106}

\bibitem[{{Cheung} {et~al.}(2015){Cheung}, {Boerner}, {Schrijver}, {Testa},
  {Chen}, {Peter}, \& {Malanushenko}}]{2015ApJ...807..143C}
{Cheung}, M. C.~M., {Boerner}, P., {Schrijver}, C.~J., {et~al.} 2015, \apj,
  807, 143, \dodoi{10.1088/0004-637X/807/2/143}

\bibitem[{{Cheung} {et~al.}(2019){Cheung}, {De Pontieu},
  {Mart{\'\i}nez-Sykora}, {Testa}, {Winebarger}, {Daw}, {Hansteen}, {Antolin},
  {Tarbell}, {Wuelser}, {Young}, \& {MUSE Team}}]{Cheung2019}
{Cheung}, M. C.~M., {De Pontieu}, B., {Mart{\'\i}nez-Sykora}, J., {et~al.}
  2019, \apj, 882, 13, \dodoi{10.3847/1538-4357/ab263d}

\bibitem[{{Culhane} {et~al.}(1991){Culhane}, {Hiei}, {Doschek}, {Cruise},
  {Ogawara}, {Uchida}, {Bentley}, {Brown}, {Lang}, {Watanabe}, {Bowles},
  {Deslattes}, {Feldman}, {Fludra}, {Guttridge}, {Henins}, {Lapington},
  {Magraw}, {Mariska}, {Payne}, {Phillips}, {Sheather}, {Slater}, {Tanaka},
  {Towndrow}, {Trow}, \& {Yamaguchi}}]{culhane1991}
{Culhane}, J.~L., {Hiei}, E., {Doschek}, G.~A., {et~al.} 1991, \solphys, 136,
  89, \dodoi{10.1007/BF00151696}

\bibitem[{{Dere} {et~al.}(2023){Dere}, {Del Zanna}, {Young}, \&
  {Landi}}]{Dere2023}
{Dere}, K.~P., {Del Zanna}, G., {Young}, P.~R., \& {Landi}, E. 2023, \apjs,
  268, 52, \dodoi{10.3847/1538-4365/acec79}

\bibitem[{{Feldman} {et~al.}(1992){Feldman}, {Mandelbaum}, {Seely}, {Doschek},
  \& {Gursky}}]{feldman1992}
{Feldman}, U., {Mandelbaum}, P., {Seely}, J.~F., {Doschek}, G.~A., \& {Gursky},
  H. 1992, \apjs, 81, 387.
\newblock
  \url{http://adsabs.harvard.edu/cgi-bin/nph-bib_query?bibcode=1992ApJS...81..387F&db_key=AST}

\bibitem[{{Golub} {et~al.}(2004){Golub}, {Deluca}, {Sette}, \&
  {Weber}}]{golub2004}
{Golub}, L., {Deluca}, E.~E., {Sette}, A., \& {Weber}, M. 2004, in Astronomical
  Society of the Pacific Conference Series, Vol. 325, The Solar-B Mission and
  the Forefront of Solar Physics, ed. {T.~Sakurai \& T.~Sekii}, 217--+

\bibitem[{{Golub} {et~al.}(2007){Golub}, {Deluca}, {Austin}, {Bookbinder},
  {Caldwell}, {Cheimets}, {Cirtain}, {Cosmo}, {Reid}, {Sette}, {Weber},
  {Sakao}, {Kano}, {Shibasaki}, {Hara}, {Tsuneta}, {Kumagai}, {Tamura},
  {Shimojo}, {McCracken}, {Carpenter}, {Haight}, {Siler}, {Wright}, {Tucker},
  {Rutledge}, {Barbera}, {Peres}, \& {Varisco}}]{golub2007}
{Golub}, L., {Deluca}, E., {Austin}, G., {et~al.} 2007, \solphys, 243, 63,
  \dodoi{10.1007/s11207-007-0182-1}

\bibitem[{{Guennou} {et~al.}(2013){Guennou}, {Auch{\`e}re}, {Klimchuk},
  {Bocchialini}, \& {Parenti}}]{guennou2013}
{Guennou}, C., {Auch{\`e}re}, F., {Klimchuk}, J.~A., {Bocchialini}, K., \&
  {Parenti}, S. 2013, \apj, 774, 31, \dodoi{10.1088/0004-637X/774/1/31}

\bibitem[{{Hannah} \& {Kontar}(2012)}]{hannah2012}
{Hannah}, I.~G., \& {Kontar}, E.~P. 2012, \aap, 539, A146,
  \dodoi{10.1051/0004-6361/201117576}

\bibitem[{{Kashyap} \& {Drake}(1998)}]{kashyap1998}
{Kashyap}, V., \& {Drake}, J.~J. 1998, \apj, 503, 450, \dodoi{10.1086/305964}

\bibitem[{{Kimble}(2011)}]{Kimble2011PhD}
{Kimble}, J. 2011, PhD thesis, University of Memphis

\bibitem[{{Kimble} \& {Schmelz}(2011)}]{kimble2011}
{Kimble}, J., \& {Schmelz}, J.~T. 2011, in American Astronomical Society
  Meeting Abstracts, Vol. 218, American Astronomical Society Meeting Abstracts
  \#218, 224.21

\bibitem[{{Lemen} {et~al.}(2011){Lemen}, {Title}, {Akin}, {Boerner}, {Chou},
  {Drake}, {Duncan}, {Edwards}, {Friedlaender}, {Heyman}, {Hurlburt}, {Katz},
  {Kushner}, {Levay}, {Lindgren}, {Mathur}, {McFeaters}, {Mitchell}, {Rehse},
  {Schrijver}, {Springer}, {Stern}, {Tarbell}, {Wuelser}, {Wolfson}, {Yanari},
  {Bookbinder}, {Cheimets}, {Caldwell}, {Deluca}, {Gates}, {Golub}, {Park},
  {Podgorski}, {Bush}, {Scherrer}, {Gummin}, {Smith}, {Auker}, {Jerram},
  {Pool}, {Soufli}, {Windt}, {Beardsley}, {Clapp}, {Lang}, \&
  {Waltham}}]{lemen2011}
{Lemen}, J.~R., {Title}, A.~M., {Akin}, D.~J., {et~al.} 2011, \solphys, 115,
  \dodoi{10.1007/s11207-011-9776-8}

\bibitem[{{Massa} {et~al.}(2023){Massa}, {Emslie}, {Hannah}, \&
  {Kontar}}]{Massa2023}
{Massa}, P., {Emslie}, A.~G., {Hannah}, I.~G., \& {Kontar}, E.~P. 2023, \aap,
  672, A120, \dodoi{10.1051/0004-6361/202345883}

\bibitem[{{O'Dwyer} {et~al.}(2014){O'Dwyer}, {Del Zanna}, \&
  {Mason}}]{odwyer2014}
{O'Dwyer}, B., {Del Zanna}, G., \& {Mason}, H.~E. 2014, \aap, 561, A20,
  \dodoi{10.1051/0004-6361/201016346}

\bibitem[{{O'Dwyer} {et~al.}(2010){O'Dwyer}, {Del Zanna}, {Mason}, {Weber}, \&
  {Tripathi}}]{odwyer2010}
{O'Dwyer}, B., {Del Zanna}, G., {Mason}, H.~E., {Weber}, M.~A., \& {Tripathi},
  D. 2010, \aap, 521, A21, \dodoi{10.1051/0004-6361/201014872}

\bibitem[{{Plowman} {et~al.}(2013){Plowman}, {Kankelborg}, \&
  {Martens}}]{plowman2013}
{Plowman}, J., {Kankelborg}, C., \& {Martens}, P. 2013, \apj, 771, 2,
  \dodoi{10.1088/0004-637X/771/1/2}

\bibitem[{{Reep} {et~al.}(2013){Reep}, {Bradshaw}, \& {McAteer}}]{reep2013}
{Reep}, J.~W., {Bradshaw}, S.~J., \& {McAteer}, R.~T.~J. 2013, \apj, 778, 76,
  \dodoi{10.1088/0004-637X/778/1/76}

\bibitem[{{Reeves} {et~al.}(2022){Reeves}, {Seaton}, {Golub}, {Cheimets},
  {DeLuca}, {DeForest}, {Del Zanna}, {Downs}, {Karna}, {Madsen}, {Moore},
  {Rivera}, {Samra}, {Savage}, {West}, \& {Winebarger}}]{kathy2022}
{Reeves}, K., {Seaton}, D.~B., {Golub}, L., {et~al.} 2022, in AGU Fall Meeting
  Abstracts, Vol. 2022, SH23A--06

\bibitem[{{Savage} {et~al.}(2023){Savage}, {Winebarger}, {Kobayashi},
  {Athiray}, {Beabout}, {Golub}, {Walsh}, {Beabout}, {Bradshaw}, {Bruccoleri},
  {Champey}, {Cheimets}, {Cirtain}, {DeLuca}, {Del Zanna}, {Dud{\'\i}k},
  {Guillory}, {Haight}, {Heilmann}, {Hertz}, {Hogue}, {Kegley},
  {Kolodziejczak}, {Madsen}, {Mason}, {McKenzie}, {Ranganathan}, {Reeves},
  {Robertson}, {Schattenburg}, {Scholvin}, {Siler}, {Testa}, {Vigil}, {Warren},
  {Watkinson}, {Weddendorf}, \& {Wright}}]{savage2022}
{Savage}, S.~L., {Winebarger}, A.~R., {Kobayashi}, K., {et~al.} 2023, \apj,
  945, 105, \dodoi{10.3847/1538-4357/acbb58}

\bibitem[{Su {et~al.}(2018)Su, Veronig, Hannah, Cheung, Dennis, Holman, Gan, \&
  Li}]{su_2018}
Su, Y., Veronig, A.~M., Hannah, I.~G., {et~al.} 2018, The Astrophysical Journal
  Letters, 856, L17, \dodoi{10.3847/2041-8213/aab436}

\bibitem[{{Testa} {et~al.}(2011){Testa}, {Reale}, {Landi}, {DeLuca}, \&
  {Kashyap}}]{Paola2011}
{Testa}, P., {Reale}, F., {Landi}, E., {DeLuca}, E.~E., \& {Kashyap}, V. 2011,
  \apj, 728, 30, \dodoi{10.1088/0004-637X/728/1/30}

\bibitem[{{Tripathi} {et~al.}(2011){Tripathi}, {Klimchuk}, \&
  {Mason}}]{tripathi2011}
{Tripathi}, D., {Klimchuk}, J.~A., \& {Mason}, H.~E. 2011, \apj, 740, 111,
  \dodoi{10.1088/0004-637X/740/2/111}

\bibitem[{{Warren} {et~al.}(2011){Warren}, {Ugarte-Urra}, {Young}, \&
  {Stenborg}}]{warren2011a}
{Warren}, H.~P., {Ugarte-Urra}, I., {Young}, P.~R., \& {Stenborg}, G. 2011,
  \apj, 727, 58, \dodoi{10.1088/0004-637X/727/1/58}

\bibitem[{{Weber} {et~al.}(2004){Weber}, {Deluca}, {Golub}, \&
  {Sette}}]{weber2004}
{Weber}, M.~A., {Deluca}, E.~E., {Golub}, L., \& {Sette}, A.~L. 2004, in IAU
  Symposium, Vol. 223, Multi-Wavelength Investigations of Solar Activity, ed.
  {A.~V.~Stepanov, E.~E.~Benevolenskaya, \& A.~G.~Kosovichev}, 321--328,
  \dodoi{10.1017/S1743921304006088}

\bibitem[{{Winebarger} {et~al.}(2011){Winebarger}, {Schmelz}, {Warren}, {Saar},
  \& {Kashyap}}]{winebarger2011}
{Winebarger}, A.~R., {Schmelz}, J.~T., {Warren}, H.~P., {Saar}, S.~H., \&
  {Kashyap}, V.~L. 2011, \apj, 740, 2, \dodoi{10.1088/0004-637X/740/1/2}

\bibitem[{{Winebarger} {et~al.}(2012){Winebarger}, {Warren}, {Schmelz},
  {Cirtain}, {Mulu-Moore}, {Golub}, \& {Kobayashi}}]{winebarger2012}
{Winebarger}, A.~R., {Warren}, H.~P., {Schmelz}, J.~T., {et~al.} 2012, \apjl,
  746, L17, \dodoi{10.1088/2041-8205/746/2/L17}

\bibitem[{{Winebarger} {et~al.}(2019){Winebarger}, {Weber}, {Bethge}, {Downs},
  {Golub}, {DeLuca}, {Savage}, {del Zanna}, {Samra}, {Madsen}, {Ashraf}, \&
  {Carter}}]{winebarger2019}
{Winebarger}, A.~R., {Weber}, M., {Bethge}, C., {et~al.} 2019, \apj, 882, 12,
  \dodoi{10.3847/1538-4357/ab21db}

\bibitem[{Wright {et~al.}(2017)Wright, Hannah, Grefenstette, Glesener, Krucker,
  Hudson, Smith, Marsh, White, \& Kuhar}]{Wright_2017}
Wright, P.~J., Hannah, I.~G., Grefenstette, B.~W., {et~al.} 2017, The
  Astrophysical Journal, 844, 132, \dodoi{10.3847/1538-4357/aa7a59}

\bibitem[{Zou \& Hastie(2005)}]{zou2005}
Zou, H., \& Hastie, T. 2005, Journal of the Royal Statistical Society Series B:
  Statistical Methodology, 67, 301, \dodoi{10.1111/j.1467-9868.2005.00503.x}

\end{thebibliography}
\bibliographystyle{aasjournal}
\end{document}